\documentclass[submission,copyright,creativecommons]{eptcs}
\providecommand{\event}{ICE 2020} 

\usepackage[utf8]{inputenc}
\usepackage{amsmath}
\usepackage{amsthm}
\usepackage{amssymb}
\usepackage{doi}
\usepackage{graphicx}
\usepackage{multicol}
\usepackage{url}
\usepackage{breakurl}
\graphicspath{ {./automata/} }
\usepackage{float}

\theoremstyle{definition}
\newtheorem{definition}{Definition}[section]

\usepackage{listings}
\usepackage{xcolor}

\definecolor{codegreen}{rgb}{0,0.6,0}
\definecolor{codegray}{rgb}{0.5,0.5,0.5}
\definecolor{codepurple}{rgb}{0.58,0,0.82}
\definecolor{backcolour}{rgb}{0.98,0.98,0.98}
\lstdefinestyle{mystyle}{
    backgroundcolor=\color{backcolour},   
    commentstyle=\color{codegreen},
    keywordstyle=\color{blue},
    numberstyle=\color{codegray},
    stringstyle=\color{codegreen},
    basicstyle=\fontsize{9}{11} \ttfamily,
    breakatwhitespace=false,         
    breaklines=true,                 
    captionpos=b,                    
    keepspaces=true,                 
    numbers=left,                    
    numbersep=5pt,                  
    showspaces=false,                
    showstringspaces=false,
    showtabs=false,                  
    tabsize=2
}

\title{Typestates to Automata and back: a tool}

\author{André Trindade
\institute{NOVA School of Science and Technology\\
NOVA University Lisbon\\
Lisbon, Portugal}
\email{adt.trindade@campus.fct.unl.pt}
\and João Mota
\institute{NOVA School of Science and Technology\\
NOVA University Lisbon\\
Lisbon, Portugal}
\email{jd.mota@campus.fct.unl.pt}
\and António Ravara
\institute{NOVA School of Science and Technology\\
NOVA University Lisbon\\
Lisbon, Portugal}
\email{aravara@fct.unl.pt}}

\def\titlerunning{Typestates to Automata and back}
\def\authorrunning{A. Trindade, J. Mota \& A. Ravara}

\begin{document}
\maketitle

\begin{abstract}
Development of software is an iterative process. Graphical tools to represent the relevant entities and processes can be helpful. In particular, automata capture well the intended execution flow of applications, and are thus behind many formal approaches, namely behavioral types.

Typestate-oriented programming allow us to model and validate the intended protocol of applications, not only providing a top-down approach to the development of software, but also coping well with compositional development. Moreover, it provides important static guarantees like protocol fidelity and some forms of progress.

Mungo is a front-end tool for Java that associates a typestate describing the valid orders of method calls to each class, and statically checks that the code of all classes follows the prescribed order of method calls.

To assist programming with Mungo, as typestates are textual descriptions that are terms of an elaborate grammar, we developed a tool that bidirectionally converts typestates into an adequate form of automata, providing on one direction a visualization of the underlying protocol specified by the typestate, and on the reverse direction a way to get a syntactically correct typestate from the more intuitive automata representation.
\end{abstract}

\section{Introduction}
Detecting software errors and vulnerabilities is becoming increasingly important in a world where the demand for code development is soaring, often leading to incomplete specifications. Building tools to help achieve this goal is crucial as testing and manual revisions have proven to be insufficient to guarantee software correctness.

Indeed, a carefully designed test suite may detect the presence of many bugs, but it does not guarantee their absence~\cite{dijkstra1972humble}. A widespread practice is the use of programming languages with type systems~\cite{cardelli1996type}. These ensure that programs do not present errors for executing invalid operations, but the type of detected errors is quite limited. Day-to-day programmers have to deal with problems that arise from not checking the correct usage of an object. For example, a type system would prevent one from using an undefined method, but not from trying to write on a file prior to opening it.

Stateful objects are non-uniform~\cite{nierstrasz1993regular}, i.e., their methods' availability depends on their internal state. Behavioral types~\cite{huttel2016foundations} are notions of types for programming languages representing the possible behavior of an entity, such as an automaton or a state machine. These notions allow us to declare behavioral specifications capturing the availability of methods. For example, a file should first be opened (once), then it could be written (multiple times, as long as there is space), or read (provided it is not empty), and when its use is finished, it should be closed (once), and cannot be used until opened again.

Textual behavioral specifications, however, can be long and cumbersome. Furthermore, not only does the existing code not use the concept of behavioral types, but it is also quite difficult to define these notions for more elaborate programs. To deal with legacy code, automatic inference tools are needed. Our goal is to enhance the support in the definition of behavioral types in programming languages like Java. We chose the Mungo tool~\cite{mungo_article}, which associates 
behavioral specifications -- \textit{Mungo protocols} or \textit{typestates} -- to Java classes and verifies if objects are used correctly. 

Given how helpful automata are in abstracting system behavior and assisting developers in understanding the underlying relation between operations, and how useful Mungo typestates are in specifying object behavior and making sure Java code respects such specification, we developed a tool that assists programmers in the (naturally iterative) process of designing 
an application. The key idea is to support the conception and visualization of (Mungo) typestates as automata.

Specifically, our main contributions are the following.
\begin{enumerate}
    \item \textbf{An automaton model equivalent to Mungo typestates:} we defined 
    a specific automaton model to Mungo protocols; besides allowing a more detailed and accurate graphical representation of the behavioral specification, it simplifies the conversion between typestate and automaton.
    \item \textbf{A grammar for Mungo typestates:} we defined the full Mungo typestates grammar, 
    providing a formal basis for constructing Mungo protocols. This detailed definition also simplifies the task of describing terms generated by the grammar as automata, given that formal grammars generally have a corresponding abstract machine.
    \item \textbf{Bi-directional translation between typestates and automata:} we have defined two algorithms that, respectively, translate Mungo protocols into automata by following the productions of the Mungo Typestate Grammar, and produce a Mungo protocol (a term of the grammar) by following all possible execution paths of an automaton. 
    \item \textbf{Implementation and web-based tool:} lastly, we developed a functional implementation of the algorithms, as well as an interactive web-based tool, which allows developers to obtain an automaton from their Mungo protocol, and, conversely, obtain a Mungo typestate from an automaton.
\end{enumerate}

\section{How to model an entity as a Mungo protocol?}\label{motivation}
\begin{figure}[htbp]
    \centering
    \includegraphics[scale=0.5]{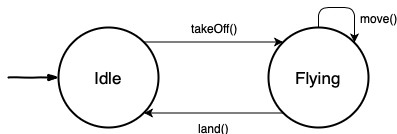}
    \caption{Automaton 1}
    \label{fig:model1}
\end{figure}

Suppose we want to build a program describing the behavior of a drone that should be able to take off, move, and land. To describe such system, we have first to understand the ordering between these actions, as it is obvious that the drone must take off prior to moving somewhere, or that it can only land if it is not yet on the ground. We can easily represent this behavior through an automaton 
(Figure \ref{fig:model1}), where each state represents those of the drone, and each transition its operations.

\begin{figure}[htbp]
    \centering
    \includegraphics[scale=0.5]{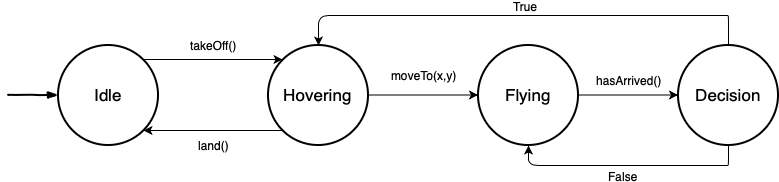}
    \caption{Automaton 2}
    \label{fig:model2}
\end{figure}

We now realize the drone should be able to move to a specified destination, but the current specification does not allow this behavior. 
We change the automaton to include a \texttt{moveTo} method instead of \texttt{move}, where \texttt{x} and \texttt{y} specify the coordinates of the desired destination (Figure \ref{fig:model2}). Additionally, we also add a method \texttt{hasArrived} that allows us to check if the drone has arrived at the specified destination.

\begin{figure}[htbp]
    \centering
    \includegraphics[scale=0.5]{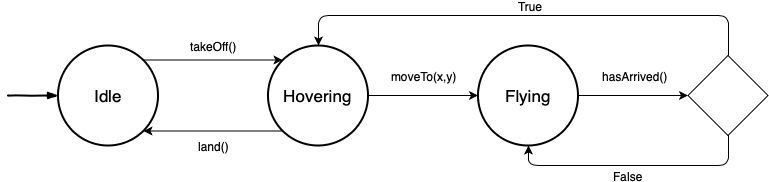}
    \caption{Automaton 3}
    \label{fig:model3}
\end{figure}

Notice that state Decision is different from the other states: transitions from this node depend on the result of method \texttt{hasArrived}, whereas transitions from other nodes are performed by executing methods. On that account, we would like to define a new kind of automaton that correctly renders this idea of states offering \textit{method-call transitions}, and states or \textit{internal-choice states} offering \textit{result-based transitions}. We call \textit{Deterministic Object Automata} (DOA) to this new type of automata, which we will be formalizing in Sec.~\ref{a:doa}. We tweak the previous automaton in order to make these changes apparent, resulting in what you see in Figure~\ref{fig:model3}.

However easy and intuitive may the representation of system behavior through state machines be, when it comes to implementation, we would be interested in having protocols that verify if, indeed, our code 
respects the behavioral model we have just designed as an automaton. To do so, one may resort to Mungo~\cite{mungo_article}, a tool used for associating automata-like specifications -- \textit{typestates} -- with Java classes. These \textit{typestate specifications} abstract the available operations on an object, by defining the sequence of permitted method-calls, which depend on the state of the object~\cite{typestate, typestate_oriented}.

Mungo protocols are defined through a grammar generating DOAs, which can then be associated to a class. It is possible to check if the associated classes are used correctly (according to the typestate), since Mungo has a typechecker. If the typestate is violated, Mungo reports the errors, otherwise we can securely compile and run the Java code using the standard Java tools.

In Listing~\ref{lst:typestate1}, we present a Mungo protocol in an attempt to define our automaton representation as a typestate that we can use to check if our code respects the intended behavior of the drone.

\newpage
\begin{lstlisting}[language=Java, label={lst:typestate1}, caption={Mungo typestate specifying the drone's behavior.}]
typestate DroneProtocol {
    Idle = { void takeOff(): Hovering }
    Hovering = { void land(): Idle, 
                 void moveTo(double, double): Flying }
    Flying = { Boolean hasArrived(): <True: Hovering, False: Flying> }
}
\end{lstlisting}

Similarly to the automaton we have previously defined, since \texttt{Idle} is the first state in our Mungo typestate, a new drone object starts in this state. From here, the only method available is \texttt{takeOff}. Calling this method changes the object's state to \texttt{Hovering}, where we can invoke method \texttt{land} and go back to state \texttt{Idle}, or invoke method \texttt{moveTo} and change to state \texttt{Flying}. In that state, executing method \texttt{hasArrived} has two possible outcomes: if the result is \texttt{True}, we change to state \texttt{Hovering}; otherwise, we stay in the same state. This mimics the idea of internal-choice states we have previously mentioned. The execution of \texttt{hasArrived} takes us to a state that evaluates the result of this method and consequently changes states according to that result.

After drafting the typestate and developing the code for the drone (see appendix~\ref{a:code1}), we now realize that this specification is quite restrictive, as it does not allow changing the drone's destination nor stop its movement mid-flight. 
We made the appropriate changes, rewriting the Mungo typestate as in Listing~\ref{lst:typestate2}.
\begin{lstlisting}[language=Java, label={lst:typestate2}, caption={Mungo typestate after changes.}]
typestate DroneProtocol {
    Idle = { void takeOff(): Hovering }
    Hovering = { void land(): Idle, 
                 void moveTo(double, double): Flying }
    Flying = { void moveTo(double, double): Flying, 
               void stop(): Hovering, 
               Boolean hasArrived(): <True: Hovering, False: Flying> }
}
\end{lstlisting}

To better understand these changes, it would be helpful 
to visualize 
the resulting automaton. By following each step of the Mungo protocol, we are able to draw 
it (Figure 4). Naturally, it could also be
obtained from the previous automaton 
by adding the corresponding transitions. The code with these modifications can be seen in appendix~\ref{a:code2}.
\begin{figure}[htbp]
    \centering
    \includegraphics[scale=0.5]{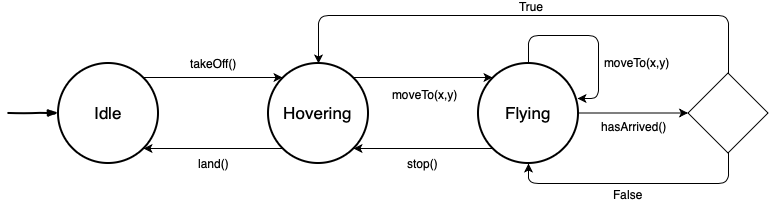}
    \caption{Automaton 4}
    \label{fig:model4}
\end{figure}

The process of describing an automaton model as a Mungo typestate, and \textit{vice versa}, can be hard and error-prone, specially when describing complex systems. In the following sections we present the work leading to the development of a tool does this transformation automatically and further allows one to directly change either the automaton or the typestate, and obtain the converse representation, by formalizing the translation between typestate and automaton. 

\section{Bi-directional conversion between Mungo Typestates and Deterministic Object Automata}
One way of looking at the problem of obtaining a Deterministic Object Automaton from a Mungo typestate (and \textit{vice versa}) is to understand similarities with typical properties regarding Formal Languages and Automata, specifically, translating a finite state automaton from a regular expression (and \textit{vice versa}). Generally, formal grammars have a corresponding state machine, for example, we can abstract regular grammars as finite automata or unrestricted grammars as Turing machines
~\cite{chomsky,hopcroft}. Therefore, formalizing a grammar for Mungo typestates naturally simplifies the task of describing these typestates as abstract machines, helping us better understand the possible behavior of an object.

Having this in mind, the first steps towards defining a conversion algorithm are the formalization of DOA, the automaton model to be equivalent to Mungo typestates, which we have highlighted in the previous section, and the formalization of a grammar for Mungo typestates. Finally, we take these two elements and, by studying their structure, define two algorithms that allow for the bi-directional translation of between both grammar and automata.

Since the end goal was to develop a tool that automates the process of translation between typestate and DOA, one of our concerns while designing such algorithms was to test that they produce the intended results
\footnote{(Mechanized) Formal proofs are work in progress.}.
For this reason, their definition was accompanied by the implementation of equivalent OCaml code~\footnote{\url{https://github.com/draexlar/compile_doa}}, a natural choice given how simple it is to represent mathematical definitions of recursive functions as functional code.

\subsection{Deterministic Object Automata}\label{a:doa}
Communicating automata have been used in the past to model binary session types~\cite{communicating_automata, heaps_and_hops}. However, the model does not fit Mungo usages perfectly, namely it does not distinguish external from internal-choice states, as clearly as DOA do. To have an automata model adequate to Mungo typestates, namely leading to an equivalence result, we defined the novel approach introduced in Sec.~\ref{motivation}. The definition of a new model allows us to have a finer control over its righteous representation of a Mungo typestate, which in the end is far easier than adapting existing models.
\medskip

\begin{definition}[\textbf{Deterministic Object Automata}]
    A DOA is an octuple $\langle S, T, M, L, s, F, D, E\rangle$ where: 
    \begin{itemize}
        \item S is a set of (external choice) states;
        \item T is a set of (internal choice) states (disjoint from S);
        \item M is a set of method identifiers;
        \item L is a set of label identifiers;
        \item $s \in S$ is the initial state;
        \item $F \subseteq S$ are the final states;
        \item $D \subseteq S \times M \times (S \cup T)$ contains the method-call transitions (external);
        \item $E \subseteq T \times L \times S$ contains all the result transitions (internal).
    \end{itemize}
\end{definition}
\newpage

\begin{definition}[\textbf{Transition function}]
    Let $\Delta = D \cup E$ be the set of transitions, with D ranged over by $\delta$ and E 
    ranged over by $\tau$. Consider $x, y \in (S \cup T)$, $a \in (M \cup L)$ and $w \in (M \cup L)^*$.
    
    The function $\Delta^* \subseteq (S \cup T) \times (M \cup L)^* \times (S \cup T)$ is inductively 
    defined by the rules:
    \begin{itemize}
        \item $\Delta^*(x, \varepsilon) = x$;
        \item $\Delta^*(x, aw) = \Delta^*(\delta(x, a), w)$, if $a \in M$;
        \item $\Delta^*(x, aw) = \Delta^*(\tau(x, a), w)$, if $a \in L$.
    \end{itemize}
\end{definition}

\subsection{Mungo Typestate Grammar}\label{grammar}
\begin{definition}[\textbf{Mungo Typestate Grammar}]
    A Mungo Typestate Grammar, $G_{Mungo}$, is a quadruple $\langle V, \Sigma, P, T \rangle$, where: 
    \begin{itemize}
        \item V is a finite set of non-terminal symbols;
        \item $\Sigma$ is a finite set of terminal symbols (alphabet), disjoint from V;
        \item P is a finite set of production rules;
        \item $T \in V$ is the start symbol.
    \end{itemize}

    \medskip
    \vspace{1mm}
    Let $N$ be a set of strings one can use to name a Mungo protocol, $K$ be a set of all possible state names, $R$ be a set of all possible data types, $Y$ be a set 
    of all possible method names, and $Z$ be a set of all possible label names. Moreover, let $name \in N$ be the name of a typestate, $state \in K$ be any valid Java identifier, $type \in R$ be any Java identifier 
    that points to a Java data type, $method \in Y$ be any valid Java identifier for a method name, and $label \in Z$ be any valid Java identifier.
    
    \medskip
    $V$ = \{ T, TB SDN, SD, S, SN, M, A, AN, W, O, ON, L, LT \}
    
    \medskip
    $\Sigma$ = \{ \texttt{typestate}, \texttt{end}, \texttt{<}, \texttt{>}, \texttt{(}, \texttt{)}, \texttt{:}, \texttt{\{}, \texttt{\}}, \texttt{,} , \texttt{=} \} $ \cup \ \{name\} \ \cup \ K \ \cup \ R \ \cup \ 
    Y \ \cup \ Z$
    
    \medskip
    \begin{table}[h!]
    \begin{tabular}{llll}
    P = \{ & \text{T $\rightarrow$ \texttt{typestate} } $name$ \text{\{TB\} }; & \text{SN $\rightarrow$ $\varepsilon$ $\mid$ , M SN };                  & \text{O $\rightarrow$ L ON };                             \\
           & \text{TB $\rightarrow$ $\varepsilon$ $\mid$ SDN TB };     & \text{M $\rightarrow$ $type$ $method$ (A) : W };                     & \text{ON $\rightarrow$ $\varepsilon$ $\mid$ , O };        \\
           & \text{SDN $\rightarrow$ $state$ = SD };                   & \text{A $\rightarrow$ $\varepsilon$ $\mid$ $type$ AN };                & \text{L $\rightarrow$ $label$: LT };                      \\
           & \text{SD $\rightarrow$ \{S\} };                         & \text{AN $\rightarrow$ $\varepsilon$ $\mid$ , $type$ AN };             & \text{LT $\rightarrow$ \texttt{end} $\mid$ $state$ $\mid$ SD }\ \} \\
           & \text{S $\rightarrow$ $\varepsilon$ $\mid$ M SN };        & \text{W $\rightarrow$ \texttt{end} $\mid$ SD $\mid$ \texttt{<} O \texttt{>} $\mid$ $state$ }; &
    \end{tabular}
    \vspace{-4mm}
    \end{table}
\end{definition}

\paragraph{Properties}
Notice that the grammar's production rules are neither left-regular nor right-regular, since their right-hand side accepts an arbitrary sequence of terminal and non-terminal symbols -- $P \subseteq V \times (V \cup \Sigma)^*$ -- as seen in the first rule of our grammar. Therefore, this is a context-free grammar.

One of our concerns while defining the grammar we present, was to ensure it to be LL(1) deterministic. This means, it is not ambiguous (produces one leftmost derivation), scans the input from left to right, and is not left-recursive. Furthermore, at each step of a derivation, there is only one applicable rule. It is simple to verify the result was achieved.

\medskip
\subsection{Translating Mungo Typestates into Deterministic Object Automata}
Now that we have formalized a grammar for Mungo protocols, we can move on to the task of defining an algorithm for translating Mungo Typestates into Deterministic Object Automata. 
To this purpose, we have defined a recursive function, $Compile$, which, given a typestate specification, infers the automaton's states and transitions by following the production rules of the Mungo Typestate Grammar, returning a DOA (as defined in Sec.~\ref{a:doa}).

\paragraph{Compile Function.}
    Please recall the sets defined in Sect.\,\ref{grammar}: $K$, the set of all possible state names; $R$, the set of all possible data types; $Y$, the set of all possible method names; and $Z$, the set of all possible label names. Let $start, next \in K$, $type \in R$, $m \in Y$, $label \in Z$, and let $G$ be the set of available states defined in the typestate\footnote{For the example given in Sect.\,\ref{motivation}, $G$ = \{\texttt{Idle}, \texttt{Hovering}, \texttt{Flying}\}. }.

    As previously noted, the $Compile$ function parses a given typestate by following the production rules of the defined Mungo Typestate Grammar. To avoid ambiguity, in cases where there is a recursive production rule, we use an apostrophe (') to distinguish symbols. For example, we denote TB~$\rightarrow$~SDN~TB as TB = SDN TB'.
    
    \medskip
    We begin with the start symbol in our grammar, T, which represents a typestate. When parsing the typestate, some symbols may be ignored since they are not relevant for constructing a DOA. For example, even though T contains more information than TB (T $\rightarrow$ \texttt{typestate} $name$ \{TB\}), we can ignore the symbols \texttt{typestate}, $name$, \{, and \}, because we are only interested in states and their respective transitions to build an automaton.
    \begin{align*}
        &Compile_G(\text{T}) = \hfill Compile_G(\text{TB})\\[6pt]
        &Compile_G(\text{TB}) = 
        \left \{
            \begin{array}{ll}
            \langle \{\texttt{end}\}, \{\}, \{\}, \{\}, \texttt{end}, \{\texttt{end}\}, \{\}, \{\}\rangle & \textit{ if } \ \text{TB} = \varepsilon \\[6pt]
            Union\big(Compile_G(\text{SDN}), \ Compile_G(\text{TB'})\big) & \textit{ if } \ \text{TB = SDN\ TB'}
            \end{array}
        \right.
    \end{align*}
    
    If the typestate's body is empty -- no states are defined -- the object is idle, and therefore the resulting automaton is characterized by having just one initial and final state: \texttt{end}, which is defined by default in any Mungo protocol. Otherwise, there is at least one defined state, thus the resulting DOA is the union (which we will define later in this section) of the automaton generated by the state SDN and the remainder of the typestate (TB').
    
    \medskip
    As an example, consider Listing~\ref{ml:ext} where we define a Mungo typestate with just two states, one initial and the final state \texttt{end}, and with only one method-call transition.
\begin{lstlisting}[language=ML, label={ml:ext}, caption={Example typestate.}]
typestate basic {
    begin = { void terminate(): end }
}
\end{lstlisting}

According to the algorithm, since the typestate's body is not empty, we would now have to compute $Union\big(Compile_G(\texttt{begin=\string{void terminate():end\string}}), \ Compile_G(\emptyset)\big)$, with $G = \{\texttt{begin}\}$. Since the typestate's body only has one state definition, the remainder of the typestate is empty. For this reason, we already know that the final union will be with an automaton with just one initial and final state, \texttt{end}: $\big\langle \{\texttt{end}\}, \{\}, \{\}, \{\}, \texttt{end}, \{\texttt{end}\}, \{\}, \{\}\big\rangle$, resulting from the computation of $Compile_G(\emptyset)$.
    
    \medskip
    Moving on with the definition of \textit{Compile}, when translating a state, if the state's body is empty (S~$= \varepsilon$) -- no transitions to other states are defined -- the resulting DOA only has one state, which is both initial and final. Otherwise, it is necessary to translate the state's transitions, as defined below.
    \begin{align*}
        &Compile_G(\text{SDN}) = Compile_G(start, \text{SD}) = Compile_G(start, \text{S}) \\[6pt]
        &Compile_G(start, \text{S}) = 
        \left \{
            \begin{array}{ll}
            \langle \{start\}, \{\}, \{\}, \{\}, start, \{start\}, \{\}, \{\}\rangle & \textit{ if } \text{S} = \varepsilon \\[6pt]
            Compile_G(start, \text{M, SN}) & \textit{ if } \text{S = M\ SN}
            \end{array}
        \right.
    \end{align*}
    
    Notice the use of $start$ instead of the name $state$ used in the grammar definition. Further on this document, we will see why this change is helpful but, since $start \in K$, it plays the same role in this instance, being $start$ the name of the current state.
    
    Since the state definition from our example is not empty, the next step in the computation of our example would be to compute the result of $Compile_G(\texttt{begin=\string{void terminate():end\string}})$, thus, according to the new step of the algorithm, we now have \[Union\big(Compile_G(\texttt{begin},\ \texttt{void terminate()},\ \texttt{end}), \ \big\langle \{\texttt{end}\}, \{\}, \{\}, \{\}, \texttt{end}, \{\texttt{end}\}, \{\}, \{\}\big\rangle\big).\]

    Continuing where we left off, in the instance of a non-empty state definition, we consider the event of only one transition being allowed -- only one method is defined in the current state (SN = $\varepsilon$) -- and the event of multiple transitions being allowed (SN = , M' SN'). In the latter, the resulting DOA is the union of the automaton generated by the first method transition (M) of the current state, and the one generated by the remainder method transitions (M' SN'). 
    \begin{align*}
        Compile_G(start, \text{M, SN}) =
        \left \{
            \begin{array}{ll}
            Compile_G(start, \text{M}) & \textit{ if } \text{SN} = \varepsilon \\[6pt]
            Union\big(Compile_D(start, \text{M}), \\\qquad\quad Compile_G(start, \text{M', SN'})\big) & \textit{ if } \text{SN = , M' SN'}
            \end{array}
        \right.
    \end{align*}
    
    To compile a method transition, we need only focus on the current state, the method allowing the transition, and the resulting state, ignoring all other symbols:
    \begin{align*}
        &Compile_G(start, \text{M}) = Compile_G(start, type\ m(\text{A}), \text{W}) \\[6pt]
        &Compile_G(start, type\ m(\text{A}), \text{W}) = \\
        & \quad
        \left \{
            \begin{array}{ll}
            \big\langle \{start, \mathtt{end}\}, \{\}, \{type \ m(\text{A})\}, \{\}, start, \{\mathtt{end}\}, \{\delta(start, \ type \ m(\text{A})) = \mathtt{end}\}, \{\}\big\rangle &\\\quad\qquad\qquad\qquad\qquad\qquad\qquad\qquad\qquad\qquad\qquad\qquad\qquad \textit{ if } \text{W} = \mathtt{end} \ or \text{ W = \texttt{\string{\string}}}\\[6pt]
            \big\langle \{start, next\}, \{\}, \{type \ m(\text{A})\}, \{\}, start, \{\}, \{\delta(start, \ type \ m(\text{A})) = next\}, \{\}\big\rangle &\\\quad\qquad\qquad\qquad\qquad\qquad\qquad\qquad\qquad\qquad\qquad\qquad\qquad \textit{ if } \text{W} = next \\[6pt]
            Union\big(\ Compile_{G \cup \{inner\}}(start,\ type\ m(\text{A}),\ inner), Compile_{G \cup \{inner\}}(inner,\ \text{SD})\ \big) &\\\quad\qquad\qquad\qquad\qquad\qquad\qquad\qquad\qquad\qquad\qquad\qquad\qquad \textit{ if } \text{W = SD},\ inner \notin G,\ inner \in K \\[6pt]
            Compile_G(start,\ type\ m(\text{A}),\ \text{O})
            \qquad\qquad\qquad\qquad\qquad\qquad\textit{ if } \text{W} = \ \texttt{<}\text{ O }\texttt{>}
            \end{array}
        \right.
    \end{align*}
    
    Once again, notice how we use $start$ instead of $state$, but also $m$ instead of $method$ and $next$ instead of $state$ which, respectively, stand for the current state, the method allowing the transition, and the resulting state. The use of $state$ would force us to work with $state =\{ type \ method$(A) : $state \}$, which is ambiguous, hence, the use of $start$ and $next$, both elements of $K$. Bear in mind that we are not concerned with which or how many arguments a method accepts, and thus we assume A is well defined.
    
    When the resulting state is the \texttt{end} state -- which can be denoted by \texttt{end} or \texttt{\string{\string}} in Mungo protocols -- the resulting automaton has: two external-choice states, $start$ and \texttt{end}; one method, $m$; an initial state, $start$; a final state, \texttt{end}; and a (\textit{method-call}) transition from $start$ to \texttt{end}, through $m$. As you can see, we follow the same reasoning when the resulting state is $next$.
    
    Mungo protocols also allow states to be defined inside other states, similar to an inlining. These ``inner'' states do not have names, so we need to assign them one. The assigned name must be unique, and thus, before assigning it, one must first check if the name is already in use, by checking if it is in $G$~\footnote{Recall that $G$ is the set of defined states in the typestate.}. We will be referring to the assigned name as $inner$ and we may see it as an element of $K$. Now, we need to update $G$ to include the newly defined inner state, so that, in the future, it will not be possible to create a state with that same name. Consequently, the generated automaton of a transition to an inner state (W = SD), is the union of ``compiling'' the current state with a resulting state $inner$, along with the resulting automaton of compiling the inner state.
    
    Lastly, Mungo protocols allow a state to transition to an \textit{internal-choice state} with a set of options (W~=~\texttt{<}~O~\texttt{>}), through a method $m$.
    
    \medskip
    According to the last few steps, we can understand that, from where we left off our example, since our state has only one defined transition, and this transition leads to state \texttt{end}, our computation results in solving: 
    \vspace{-2mm}
    \begin{align*}
        Union\Big(\ &\big\langle \{\texttt{begin}, \mathtt{end}\}, \{\}, \{\texttt{void terminate()}\}, \{\}, \texttt{begin}, \{\mathtt{end}\},\\
        &\ \ \{\delta(\texttt{begin}, \texttt{ void terminate()}) = \mathtt{end}\}, \{\}\big\rangle,\ \big\langle \{\texttt{end}\}, \{\}, \{\}, \{\}, \texttt{end}, \{\texttt{end}\}, \{\}, \{\}\big\rangle\ \Big).
    \end{align*}
    
    Continuing with the definition of \textit{Compile}, we have still to define the translation of transitions to internal-choice states.
    \begin{align*}
        &Compile_G(start,\ type\ m(\text{A}),\ \text{O}) = \\
        &\qquad
        Union\Big(\big\langle \{start\}, \{choice\}, \{type \ m(\text{A})\}, \{\}, start, \{\},\ \{\delta(start, \ type \ m(\text{A})) = choice\}, \{\}\big\rangle,\\ 
        &\ \ \quad\qquad\qquad Compile_{G \cup \{choice\}}(choice, \text{ O})\Big),\ 
        \text{where } \ choice \notin G, \ choice \in K
    \end{align*}
    
    To translate a transition to an internal-choice state, we must first assign a name to that state, which we will be referring to as $choice$, and we may see it as an element of $K$. Although $S$ (the set of external-choice states) and $T$ (the set of internal-choice states) are disjoint, $choice$ should be unique to avoid ambiguity. So, similarly to what was done for the inner states, we check if $choice$ is in $G$, and then update this set. The resulting DOA is the union of an automaton with: one external-choice state, $start$; one internal-choice state, $choice$; a method, $m$; an initial state; and an (method-call) transition from $start$ to $choice$; along with the resulting automaton of translating the set of options (O) offered by $choice$.
    \begin{align*}
        &Compile_G(choice, \text{ O}) = Compile_G(choice, \text{L, ON})\\[6pt]
        &Compile_G(choice, \text{L, ON}) = 
        \left \{
            \begin{array}{ll}
            Compile_G(choice,\ \text{L}) & \textit{ if } \text{ON} = \varepsilon \\[6pt]
            Union(Compile_G(choice,\ \text{L}), \ Compile_G(choice,\ \text{O})) & \textit{ if } \text{ON = , O}
            \end{array}
        \right.
    \end{align*}
    
    Notice that, our grammar does not accept an internal-choice state -- a decision state -- without options. Hence, we consider the instances of only one option being defined in the current decision state (ON~=~$\varepsilon$), and the instance of having multiple options (ON = , O). We then “compile” the decision state with its first option (or only option). By performing its union with the automaton generated by the the remainder options in $choice$, we satisfy the latter instance.
    
    \medskip
    To compile an option, we focus on the current internal-choice state, $choice$, the option's label identifier, $label$, and the option's resulting (external-choice) state for the label in consideration, LT:
    \begin{align*}
        &Compile_G(choice, \ \text{L}) = Compile_G(choice,\ label, \text{ LT}) \\[6pt]
        &Compile_G(choice,\ label, \text{ LT}) = \\
        &\quad
        \left \{
            \begin{array}{ll}
            \langle \{\mathtt{end}\}, \{choice\}, \{\}, \{label\}, \epsilon, \{\mathtt{end}\}, \{\}, \{\tau(choice, label) = \mathtt{end}\}\rangle \quad \textit{if } \text{LT} = \texttt{end} \ or \text{ LT = } \texttt{\string{\string}}\\[6pt]
            \langle \{next\}, \{choice\}, \{\}, \{label\}, \epsilon, \{\}, \{\}, \{\tau(choice, label) = next\}\rangle \ \quad\quad
            \textit{if } \text{LT} = next\\[6pt]
            Union\big(Compile_{G \cup \{inner\}}(choice,\ label,\ inner), \ Compile_{G \cup \{inner\}}(inner,\ \text{SD})\big) \\ \ \ \ \qquad\qquad\qquad\qquad\qquad\qquad\qquad\qquad\qquad\qquad\qquad\qquad\
            \textit{if } \text{LT} = \text{SD}, \ inner \notin G,\ inner \in K
            \end{array}
        \right.
    \end{align*}
    
    Similarly to what was done for translating method-call transitions, when our resulting state is the \texttt{end} state, the resulting automaton has: one external-choice state, \texttt{end}; one internal-choice state, $choice$; one label; a final state; and a (\textit{result-based}) transition from $choice$ to \texttt{end}, through $label$. Otherwise, when the resulting state is $next$, the resulting automaton has: one external-choice state, $next$; one internal-choice state, $choice$; one label; and a (result-based) transition from $choice$ to $next$, through $label$.
    
    Mungo typestates also allow states to be defined inside internal-choice states (LT = SD). Therefore, the resulting automaton is the union of translating the current internal-choice state with a resulting state $inner$ ($inner \in K$ and $inner \notin G$) through $label$, along with the resulting automaton of compiling the inner state. Recall that we then need to update $G$ to include the new state $inner$, so that, in the future, no other state can have the same name.

\paragraph{Union of Deterministic Object Automata.}
At many steps in our $Compile$ function we made use of a $Union$ function. This is because simply performing the union ($\cup$) of two Deterministic Object Automata is not right. Although it would be valid for almost every set in our octuple, notice that, in the definition of DOA, there is only one initial state, which is assumed to be the first state defined in a Mungo typestate. Addressing this issue, a function $Union(a, b)$ is defined.

\medskip
\begin{definition}[\textbf{Union Function}]
    Let $a$ and $b$ be two DOA, where $a$ is the automaton with the initial state considered to be the start:
    \begin{equation*}
        Union(a, b) = \big\langle \ S_a \cup S_b,\ T_a \cup T_b,\ M_a \cup M_b,\ L_a \cup L_b,\ s_a, F_a \cup F_b,\ D_a \cup D_b,\ E_a \cup E_b \ \big\rangle
    \end{equation*}
\end{definition}

\noindent
Now that the \textit{Union} function is defined, we can now complete the last computation from our example, which we recall bellow.
\begin{align*}
    Union\Big(\ &\big\langle \{\texttt{begin}, \mathtt{end}\}, \{\}, \{\texttt{void terminate()}\}, \{\}, \texttt{begin}, \{\mathtt{end}\},\\ 
    &\ \ \{\delta(\texttt{begin}, \texttt{ void terminate()}) = \mathtt{end}\}, \{\}\big\rangle,\ \big\langle \{\texttt{end}\}, \{\}, \{\}, \{\}, \texttt{end}, \{\texttt{end}\}, \{\}, \{\}\big\rangle\ \Big)
\end{align*}

Notice that, in this particular case, the first automaton already has a transition to \texttt{end}, therefore the resulting sets will be equal to those of this DOA. Furthermore, by definition, the initial state of the first automaton is the initial state of the resulting automaton, thus we obtain the DOA: $\big\langle \{\texttt{begin}, \mathtt{end}\}, \{\},$ $\{\texttt{void terminate()}\}, \{\}, \texttt{begin}, \{\mathtt{end}\}, \{\delta(\texttt{begin}, \texttt{ void terminate()}) = \mathtt{end}\}, \{\}\big\rangle$.

\subsection{Translating Deterministic Object Automata into Mungo Typestates}
We now define the converse translation. By describing the behavior of an entity as a state machine, one should be able to infer its corresponding Mungo protocol. This would be helpful since describing an automaton can be easier and more intuitive than writing a typestate specification.

In this section, we propose a method for translating Deterministic Object Automata into Mungo typestates by defining a function, $Decompile$, that, given a DOA, infers a corresponding Mungo typestate by following the automaton's transitions.

\paragraph{Decompile function.}
    Let $name$ be the name one wants to give the resulting Mungo typestate, and $doa = \langle S, T, M, L, s, F, D, E\rangle$ be a Deterministic Object Automaton. And let $G$ be the set of already defined states in the typestate. Initially, $G = \{\texttt{end}\}$ given that the \texttt{end} state is predefined for every Mungo typestate.
    The $Decompile$ function returns a string corresponding to the Mungo protocol described by \textit{doa}.

    We start by giving the typestate's name and the automaton \textit{doa} representing the typestate as arguments of our \textit{Decompile} function, in order to obtain the protocol's header, followed by the definition of its body.
    \begin{align*}
        &Decompile(name, doa) = \texttt{typestate} \ name \ \{ \ Decompile_G(doa) \ \}\\[6pt]
        &Decompile_G(doa) =
        \left \{
            \begin{array}{ll}
                \varepsilon \qquad\qquad\qquad\qquad\qquad\qquad\qquad\qquad\qquad\qquad\qquad\
                \textit{ if} \ S = \emptyset \ or \ S = \{\texttt{end}\} \\[6pt]
                s = \{ \ Decompile_{G \cup \{s\}}(s, doa, A) \ \} \\\quad
                Decompile_{G \cup \{s\}}\big(\langle S \setminus \{s\}, T, M, L, n, F, D \setminus A, E \rangle\big),\\\qquad\ \
                \text{with } A = \{ \delta(x, y)=z \in D \mid x=s \} \text{ and } n \in S \setminus \{s, \text{end}\}
                \qquad\quad \textit{ otherwise }
            \end{array}
        \right.
    \end{align*}
    
    Looking at the Mungo Typestate Grammar definition, we see that the typestate's body can be empty (TB $\rightarrow$ $\varepsilon$ $\mid$ SDN TB). Therefore, if the set of external-choice states, $S$, of the given DOA is empty or its only element is \texttt{end}, there are no (more) states to be defined in the typestate, and this instance of $Decompile$ returns the empty string. Otherwise, we define state $s$ in the typestate. 
    
    Recall that the first state defined in a Mungo protocol is the initial state of the typestate and, naturally, the initial state of the $doa$, $s$. For this instance, we also need to define the remaining states of the given automaton. Thus, we make the recursive call of $Decompile$ with a DOA where: the set of external-choice states, $S$, no longer includes $s$; its initial state can be any state in $S \setminus \{s, \text{end}\}$; and the set of method-call transitions, $D$, does not include any transition where the initial state is $s$. Lastly, it is important that we update the set of defined states in the protocol, $G$, to include the newly defined state $s$.
    
    \medskip
    As an example, consider a DOA with just two states, one initial and one final with only one method-call transition:
    \[d = \big\langle \{begin\},\ \emptyset,\ \{\texttt{void terminate()}\},\ \emptyset,\ begin,\ \{\texttt{end}\},\ \{(begin, \texttt{ void terminate()}, \texttt{ end})\},\ \emptyset \big \rangle.\]

According to the first steps of \textit{Decompile}, if we called $Decompile(\texttt{basic}, d)$, we would have $s = begin$, $A=\{(begin, \texttt{ void terminate()}, \texttt{ end})\}$, and $G=\{\texttt{end},\ begin\}$, given that $begin$ is the initial state and the only transition beginning in $begin$ is that of calling method \texttt{terminate}.

\medskip
Continuing with the formalization of \textit{Decompile}, the body of an external-choice state is the state's method-call transitions. For this reason, besides the current state, $c$, and $doa$, we also receive the set of all the current state's method-call transitions, $A$, as follows.

\begin{align*}
    &Decompile_G(c, doa, A) =\\&\qquad
    \left \{
        \begin{array}{ll}
            \varepsilon & \textit{ if } A = \emptyset \\[6pt]
            m:\ Decompile_G(n, doa) & \textit{ if } A = \{\delta(c, m) = n\},\ m \in M \\[6pt]
            m:\ Decompile_G(n, doa), \\\quad\ \ \ Decompile_G\big(c, doa, A \setminus \{\delta(c, m) = n\}\big) & \textit{ if } \#A \neq 1,\ \{\delta(c, m) = n\} \subset A,\ m \in M
        \end{array}
    \right.
\end{align*}
    
If the set of method-call transitions from $c$ is empty, there are no (more) transitions to be defined for the current state. If the set of method-call transitions from $c$ only has one element, then we need to define the transition relative to that element. Taking a look at our grammar definition, we define a transition by writing the method allowing it, followed by colon (:) and the resulting (external-choice) state or a series of choices (internal-choice state). 

Otherwise, the set of method-call transitions has multiple elements. We start by choosing one of its elements and defining it, similarly to what was explained in the prior instance, followed by a comma (,) and the definition of the remaining transitions allowed in the current state, $c$. This is done by recursively calling $Decompile$ with a set of method-call transitions excluding the one we have just defined.
    
To finish defining a transition, we need to decide whether the state we are transitioning to, $n$, is external or internal.
    \begin{align*}
        Decompile_G(n, doa) =
        \left \{
            \begin{array}{ll}
                n & \textit{ if } n \in S \cup G \\[6pt]
                \texttt{< } Decompile_G\big(n, doa,\{\tau(x, y) =z \in E \mid x = n\}\big) \texttt{ >} & \textit{ if } n \in T
            \end{array}
        \right.
    \end{align*}
    
    To do so, we need only check if $n$ is in $S$, the set of external-choice states, or in $G$, the set of already defined states; or otherwise check if $n$ is in $T$, the set of internal-choice states. In which case, we start by writing \texttt{<}, which identifies the choice; call $Decompile$ with the state we are transitioning to, $n$, the automaton $doa$, and the set of all result transitions starting in $n$, which we will be calling $B$; followed by closing the choice with \texttt{>}.
    \begin{align*}
        &Decompile_G(c, doa, B) =\\&\quad
        \left \{
            \begin{array}{ll}
                l: n & \textit{ if } B = \{\tau(c, l) = n\}, \ l \in L,\ n \in S \cup G \\[6pt]
                l: n,\ Decompile_G\big(c, doa, B \setminus \{\tau(c, l) = n\}\big) & \textit{ if } \#B > 1,\ \{\tau(c, l) = n\} \subset B,\ l \in L, \ n \in S \cup G
            \end{array}
        \right.
    \end{align*}
    
    In the Mungo Typestate Grammar definition, you can see there must be at least one choice in an internal-choice state (O $\rightarrow$ L ON). Therefore, we consider the instance of $B$ (the set of result transitions starting in $c$) having only one element, and the instance of $B$ having multiple elements. 
    
    Taking a look at the grammar definition, in a Mungo typestate, we define a result transition by writing a label, $l$, followed by colon (:), and the resulting external-choice state, $n$. Since $n$ is an external-choice state, it must be an element of $S$ or an element of $G$. When having multiple result transitions for $c$, we start by choosing one of its elements and defining it, as explained above, followed by a comma (,), and the definition of the remaining result transitions allowed in the current state. This is done by recursively calling $Decompile$ with a set of result transitions excluding the one we have just defined.

    \medskip
    From these lasts few steps of \textit{Decompile} we can see that, given that set $A$ is the singleton $A=\{(begin, \texttt{ void terminate()}, \texttt{ end})\}$ and that, indeed, $\texttt{end} \in S \cup G$, the resulting typestate from our previous example is the following. 
\begin{lstlisting}[language=ML, label={ml:exa3}, caption={Typestate obtained from $d$.}]
typestate basic {
    begin = { void terminate(): end }
}
\end{lstlisting}
\vspace{-4mm}

\section{The Web Tool}
\vspace{-2mm}
The work presented in the previous section gave us the formal basis to what we sought out to do from the beginning: build a tool~\footnote{\url{http://typestate-editor.github.io/}} that helps developers describe behavioral specifications by automating the process of converting Mungo typestates into automata and \textit{vice versa}.
This tool allows any user to write a Mungo protocol and preview a visual representation of the corresponding automaton.

The implementation was done using \texttt{TypeScript}~\footnote{\url{https://www.typescriptlang.org/}}, a static type checker for JavaScript. Although we tried to stick to the functional approach, discussed in the previous section, as to follow the \textit{Compile} and \textit{Decompile} function definitions, some parts of the implementation are imperative in nature, since they work with mutable structures for performance sake. Lastly, to help us with handling the data and the graphical representation of typestates as automata, we used the \texttt{vis.js}~\footnote{\url{https://visjs.org/}} library.

\begin{figure}[H]
    \centering
    \includegraphics[scale=0.45]{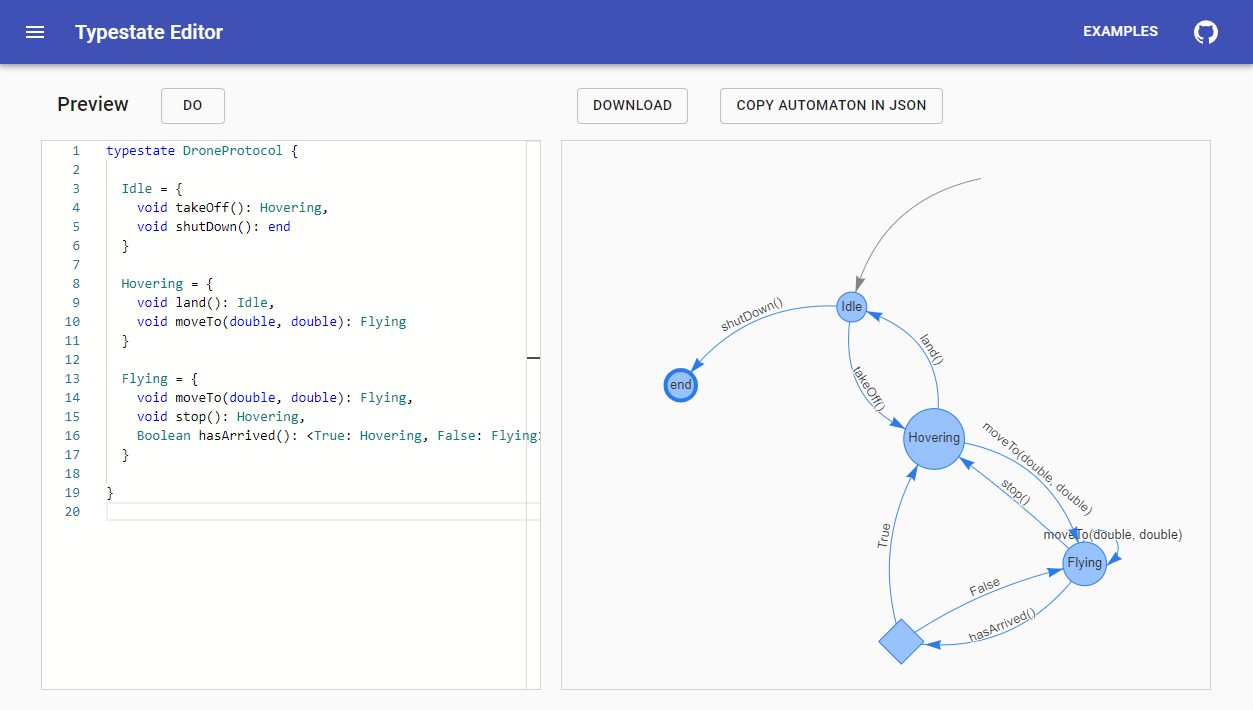}
    \vspace{-2mm}
    \caption{Web tool interface: Mungo typestate to DOA.}
    \label{fig:typestate_to_preview}
\end{figure}

Figure~\ref{fig:typestate_to_preview} presents the tool's interface when converting a Mungo protocol into an Deterministic Object Automata. On the left, one can type the typestate specification and click the ``Do'' button to produce a corresponding visual representation -- a DOA -- which can be observed on the right. Each element of the image has a meaning: the gray arrow points to the initial state; blue arrows represent transitions; external-choice states are represented as circles; and internal-choice states are represented as diamonds. Finally, one can also download the automaton as a \texttt{PNG} file or copy the automaton in \texttt{JSON} to the clipboard.



\begin{figure}[htbp]
    \centering
    \includegraphics[scale=0.45]{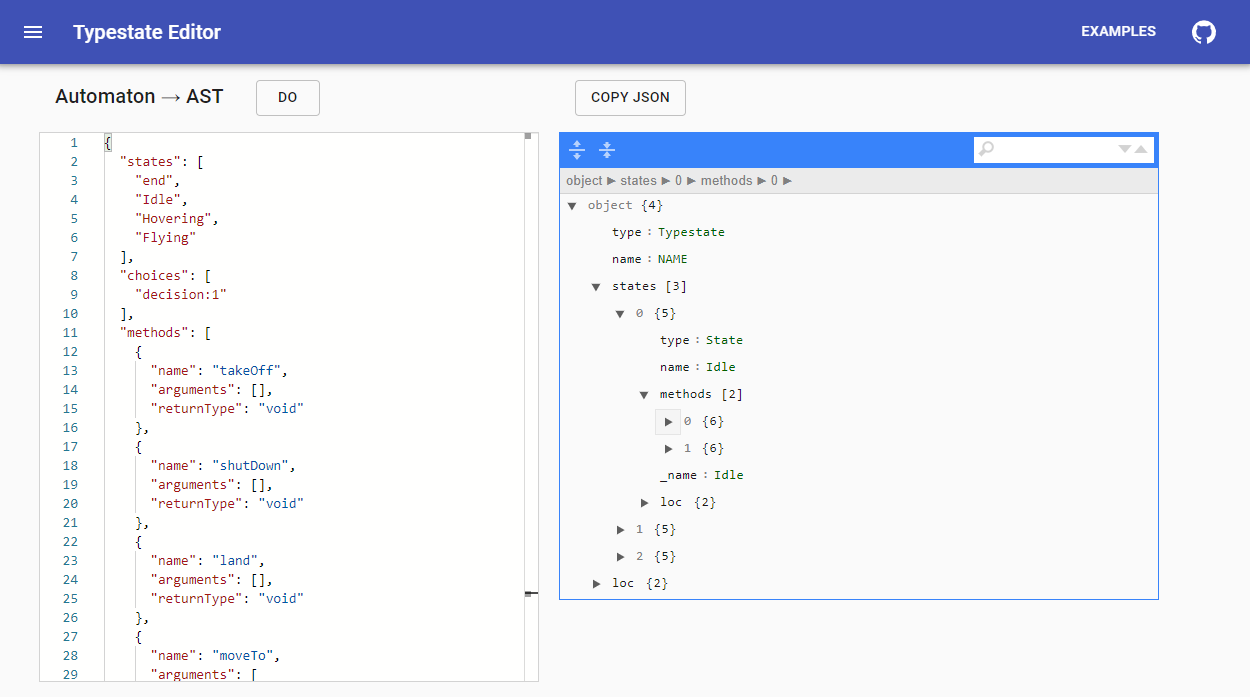}
    \vspace{-2mm}
    \caption{Web tool interface: DOA to AST.}
    \label{fig:automaton_to_ast}
\end{figure}
\begin{figure}[H]
    \centering
    \includegraphics[scale=0.45]{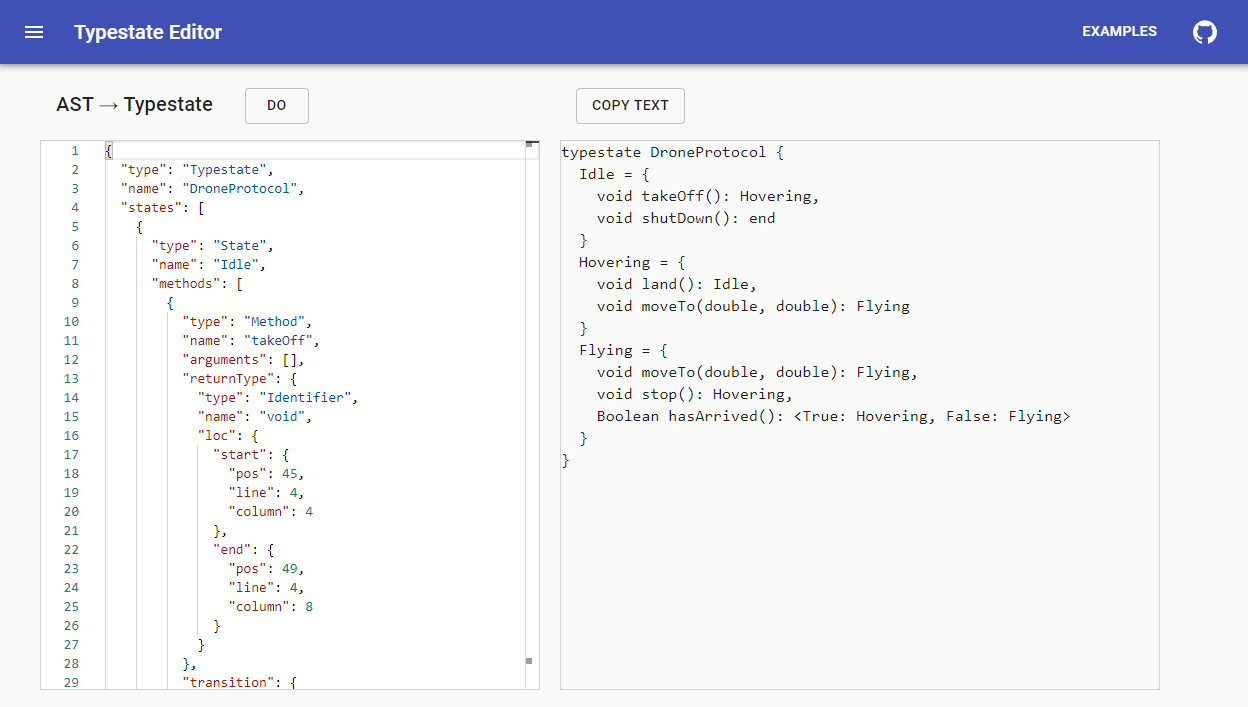}
    \vspace{-2mm}
    \caption{Web tool interface: AST to Mungo typestate.}
    \label{fig:ast_to_typestate}
\end{figure}

Additionally, the user may view the internal representations and intermediate transformations to understand how the tool works. Namely, the user may transform the typestate into an abstract syntax tree (and \textit{vice versa}), and transform an abstract syntax tree into an automaton (and \textit{vice versa}), both represented in \texttt{JSON}. Figure~\ref{fig:automaton_to_ast} shows an example of a transformation from an automaton to the abstract syntax tree, which can then be used to obtain the corresponding typestate, as illustrated in Figure~\ref{fig:ast_to_typestate}. Notice how the inverse transformation produced the same typestate initially presented in Figure~\ref{fig:typestate_to_preview}.
\vspace{-2mm}

Lastly, we would like to add that the tool also checks for errors. In the written typestates, it detects syntax errors and errors related to the structure of the typestate, such as: states with the reserved name \texttt{end}; references to states that are not defined; duplicate states, transitions (for the same state) or labels (in internal-choice states); and internal-choice states without options. When performing the converse translation, we look for errors regarding transitions to undefined states, internal-choice states without result transitions, or even transitions from internal-choice states to internal-choice states.

\section{Conclusions and Further work}
Nowadays, programming languages 
rely heavily on type systems to avoid non-trivial errors such as the execution of 
operations on invalid states, 
as trying to access a non-existent position of an array or trying to read from a file that is not yet opened. Behavioral types 
are useful methods to prevent errors resulting from applying operations in wrong order. 

In this paper we focus on 
how tools such as Mungo allow us to describe the behavior of entities through its typestates. 
By presenting a formal grammar description of Mungo protocols, we define a norm for building Mungo typestates, and by designing a new automata model that can soundly portray a Mungo typestate, we give a concrete way of visualizing the described object's behavior. The formalization of a function for translating Mungo typestates into Deterministic Object Automata, as well as a function that computes the reverse translation, grants mathematical grounds for developing tools to represent and manipulate behavioral descriptions of computational entities. Such is the web-based tool we have developed, which, through an easy to use graphical interface, allows any developer to obtain a Mungo typestate from an automaton, or even get a better sense of how their Mungo typestate is structured, by displaying an equivalent automaton graph. Allied with the Mungo tool, one can associate the protocols designed or obtained through our tool with corresponding Java class implementations, and Mungo's typechecker will verify if the code follows the revised protocols.


Nonetheless, as many have said, \textit{a work of art is never finished}, and so is true for the work herein presented. A common theme throughout this paper, and the motivation driving the use of behavioral types is, in its core, the same as for using other formal methods: the necessity of systems that ensure correct behavior. Therefore, the next step in our work is to ensure the correct specification of our algorithms, by using deductive program verification to prove that the behavior and properties we expect from them are correct. Concretely, the formal proof is based on the typical properties regarding Formal Languages and Automata: the conversion between grammar and automaton preserve the language. This is, to prove the correction of the algorithm that converts Mungo typestates into DOA is to prove that the \textit{language} of the (resulting) DOA is the same as the \textit{language} of the Mungo typestate. Similarly, to prove the correction of the translation from DOA to Mungo typestate is to prove that the \textit{language} of the (resulting) Mungo typestate is the same as the \textit{language} of the DOA. The result of this work will contribute to the development of a mechanically verified version of the tool that developers can use with the assurance of correct results. Following this step, we intend on improving the usability of the tool, by allowing the users to construct and manipulate the graphical representation of the automaton, rather than having to work with the formal, more complex, textual representation.

\bibliographystyle{eptcs}
\bibliography{bibliography.bib}

\newpage
\appendix

\section{Drone with ability to check its arrival}\label{a:code1}

\begin{lstlisting}[language=Java]
typestate DroneProtocol {
    Idle = {
        void takeOff(): Hovering,
        void shutDown(): end
    }
    Hovering = {
        void land(): Idle,
        void moveTo(double, double): Flying
    }
    Flying = {
        Boolean hasArrived(): <True: Hovering, False: Flying>
    }
}
\end{lstlisting}

\begin{lstlisting}[language=Java]
import mungo.lib.Typestate;
@Typestate("DroneProtocol")
public class Drone {
  public Drone() {}
  void takeOff() {}
  void land() {}
  void moveTo(double x, double y) {}
  void shutDown() {}
  Boolean hasArrived() {
    return Boolean.False;
  }
}
\end{lstlisting}

We noticed that the previous typestate was too restricted so, we extended it so that we can check if the drone has arrived to its destination. In the following example, after telling the drone to move, we wait in a loop until the drone arrives. We use a \texttt{do-while} loop, \texttt{break}/\texttt{continue} statements and a \texttt{switch} statement so that Mungo understands the flow of execution. To simplify the code, we introduced a \texttt{shutDown} method so that we reach the \texttt{end} state, making Mungo accept our code -- without the need to create another loop around it, similar to the first example.

\begin{lstlisting}[language=Java]
public class Main {
	public static void main(String[] args) {
        Drone drone = new Drone();
        drone.takeOff();
        drone.moveTo(20.0, 10.0);
        loop: do {
          switch(drone.hasArrived()) {
            case True:
              break loop;
            case False:
              continue loop;
          }
        } while(true);
        drone.land();
        drone.shutDown();
	}
}
\end{lstlisting}

\newpage
\section{Drone with ability to change course while flying}\label{a:code2}
\begin{lstlisting}[language=Java]
typestate DroneProtocol {
    Idle = {
        void takeOff(): Hovering,
        void shutDown(): end
    }
    Hovering = {
        void land(): Idle,
        void moveTo(double, double): Flying
    }
    Flying = {
        void moveTo(double, double): Flying,
        void stop(): Hovering,
        Boolean hasArrived(): <True: Hovering, False: Flying>
    }
}
\end{lstlisting}

This final example is very similar to the previous one. We just introduced the possibility of changing the drone's course by allowing the method \texttt{moveTo} to be called in the \texttt{Flying} state. This code is also accepted by Mungo.

\begin{lstlisting}[language=Java]
public class Main {
	public static void main(String[] args) {
        Drone drone = new Drone();
        drone.takeOff();
        drone.moveTo(20.0, 10.0);
        drone.moveTo(10.0, 20.0);
        loop: do {
          switch(drone.hasArrived()) {
            case True:
              break loop;
            case False:
              continue loop;
          }
        } while(true);
        drone.land();
        drone.shutDown();
	}
}
\end{lstlisting}


\end{document}